# The stable behavior of low thermal conductivity in 1T-sandwich structure with different components

E Zhou[1], Jing Wu[1], Chen Shen[2], Hongbin Zhang[2], and Guangzhao Qin[1,*]

[1]*State Key Laboratory of Advanced Design and Manufacturing for Vehicle Body, College of Mechanical and Vehicle Engineering, Hunan University, Changsha 410082, P. R. China*
[2]*Institut für Materialwissenschaft, Technische Universität Darmstadt, Darmstadt, 64289, Germany*

**Abstract:** Designing materials with low thermal conductivity ($\kappa$) is of demand for thermal protection, heat insulation, thermoelectricity, *etc*. In this paper, based on the *start-of-art* first-principles calculations, we propose a framework of a 1T-sandwich structure for designing materials with low $\kappa$. The 1T-sandwich structure is the same as the well-known transition metal dichalcogenide (TMD) but with light Carbon atoms in the middle plane. Using different atoms to fill the outer positions, a few novel two-dimensional materials are constructed as study cases, *i.e.,* $Mg_2C$, Janus MgBeC, $Be_2C$, and $Mo_2C$. With a systematic and comparative study, the $\kappa$ are calculated to be 3.74, 8.26, 14.80, 5.13 W/mK, respectively. The consistent values indicate the stable behavior of low $\kappa$ in the 1T-sandwich structure, being insensitive to the component. Our study would help design advanced functional materials with reliable heat transfer performance for practical applications, which reduces the influence of unavoidable impurities.

---

[*] Author to whom all correspondence should be addressed. E-Mail: gzqin@hnu.edu.cn

# 1. Introduction

Recently, green and environment-friendly new energy vehicles (NEV) have become increasingly popular. Fighting the energy crisis, the efficient use of energy, and the secondary use of waste heat play a significant role.[1,2] The emergence of advanced functional materials with low thermal conductivity ($\kappa$) greatly improve the energy conversion efficiency of engines[3], increase waste heat recovery, and play a leading role in the next generation of the automobile industry. Specifically, advanced materials with low $\kappa$ can be applied in gas-turbine, known as thermal-barrier-coatings (TBCs) materials[4], which have outstanding thermal protection features. For instance, the lower $\kappa$ of materials benefits a higher fire resistance limit, which is also useful for fire suits. Besides, it can be used as architectural insulation materials[5], saving energy and keeping warm. In addition, materials with low $\kappa$ are more promising for the application in thermoelectrics[6–9], which have made significant progress in the space industry. Thermoelectric materials with low $\kappa$ can effectively convert thermal energy from solar into electricity, which have an active effect on the performance of high-temperature equipment in the extreme environment. Thus, it can be seen that materials with low $\kappa$ will play a crucial role in future development.

Since the discovery of graphene in 2004,[10] searching novel two-dimensional (2D) materials with unique properties has sparked a boom.[11–13] To meet the increasing practical demands, many thermal regulating methods have emerged, such as composition design, structural design,[14,15] strain engineering,[16–18] electric field,[19,20] defect,[21] vacancies,[22] *etc*. For graphene-like monolayer honeycomb structure, a component replacement can realize the step-less adjustment of the $\kappa$, such as black phosphorene,[23] silicene[24] and germanene[25]. Inspired by the study on the phonon thermal properties of 2D materials published by Gu *et al.*, materials with 1T structure have relatively lower $\kappa$.[26] Moreover, benefiting from the formed resonant bonding, the $\kappa$ of 1T-sandwich structure with light Carbon (C) atoms in the middle plane can be further lowered.[27] Besides, it is worth noting that impurities of materials are common phenomena in practical applications and experimental synthesis. Impurities will lead to additional scattering of phonon transport, which have significant impacts on the thermal transport properties, makes the result unreliable,[28,29] and then affects the credibility of materials. Thus, it is of extraordinary significant to study the thermal transport properties with components replaced by different atoms.

This study systematically studied the 1T-sandwich structure with light C atoms in the middle plane. Using the component design method with different atoms to fill the outer positions, we constructed a few novel two-dimensional materials to have the low $\kappa$ as study cases, *i.e.,* $Mg_2C$, $Be_2C$, Janus MgBeC, and also $Mo_2C$, which are found to be insensitive to the composition. In the following parts, we will have a comparative analysis of thermal transport performance from the aspects of structure, phonon spectrum, mode analysis, and electronic structure. Finally, we summarized to explain the mechanisms underlying the stable behaviour of low thermal conductivity.

## 2. Computational methodology

Using the *Vienna ab initio simulation package* (VASP)[30], all the first-principles calculations are performed based on the density functional theory (DFT). The generalized gradient approximation (GGA) of Purdue-Burke-Ernzerhof (PBE)[31] functional deals with the exchange-correlation interaction. The energy cut-off was set as 1000 eV, and the threshold of energy convergence is $10^{-6}$ eV. The Monkhorst-Pack[32] $k$-points mesh of $8\times8\times1$, $10\times10\times1$, $11\times11\times1$, and $13\times13\times1$ are used for sampling the first Brillouin region of $Mg_2C$, $Be_2C$, Janus MgBeC, and $Mo_2C$, respectively. Besides, the structural optimizations would not stop unless the force convergence accuracy reached $10^{-5}$, $10^{-8}$, $10^{-7}$, and $10^{-8}$ eV/Å, respectively. Utilizing the PHONOPY[33] and thirdorder.py[34], the harmonic (2$^{nd}$) and anharmonic (3$^{rd}$) force constants were obtained. A $5\times5\times1$ super-cell and the finite displacement difference method of real space were employed in calculating the 2$^{nd}$ and 3$^{rd}$ force constant to make the results accurate. The cut-off radius is introduced in the calculation of the third-order force constant. Based on the $\kappa$ and the trace of interatomic force constant tensors versus atomic distance convergence test[35] as shown in Supplemental Fig. S1, the fifth, sixth, seventh, seventh nearest-neighbor atoms are adopted as the cut-off radius for the four materials, respectively. In addition, the Born effective charge ($Z^*$) and the dielectric tensor ($\varepsilon$) derived from density functional perturbation theory (DFPT) are added to the dynamic matrix as corrections to take into account the long-range electrostatic interactions. Based on the results above, the phonon Boltzmann transport equation (BTE) was solved by ShengBTE software package[36] with a $101 \times 101 \times 1$ **q**-point grid. Finally, the $\kappa$ and related heat transfer properties were obtained. Besides, we performed ab-initio molecular dynamics (AIMD) simulations using the NVT ensemble of 8000 time-steps with a timestep of 1 fs at 300 K to examine the thermal stability of Janus MgBeC (Supplemental Fig. S2).

## 3. Results and discussion

### 3.1 Lattice thermal conductivity

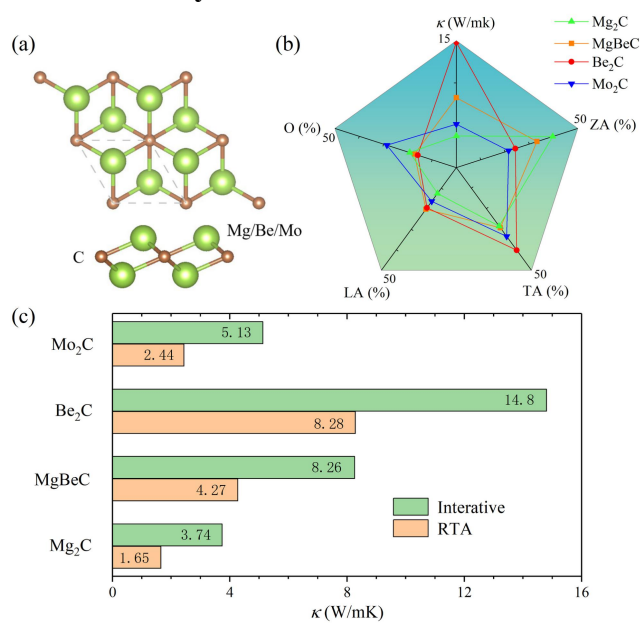

Fig. 1 (a) Top and side views of the 1T-sandwich structure with light C atoms in the middle plane. The brown and green balls represent C and Mg/Be/Mo atoms, respectively. (b) Percentage contribution of each phonon branch (ZA, TA, LA, and O) to the thermal conductivity ($\kappa$). (c) Comparison of $\kappa$ for the four different materials using the iterative method and relaxation time approximate (RTA).

As shown in Fig. 1(a), different from the transition metal chalcogenides (TMCs) with 2H structure,[37] the lighter C atoms lie in the middle plane of the 1T-sandwich structure, forming a hexa-coordinate C atoms plane[27,38]. Employing the concept of composition design, we purposely fill the top and bottom sublayers with the group IIA atoms. It is found that a class of novel single-layer materials with a 1T-structure can be formed, such as $Mg_2C$, $Be_2C$, and Janus MgBeC. In addition to the lighter atoms of the group IIA, outer-layer atoms can also be replaced by particular transition elements to make novel stable materials, such as $Mo_2C$, $Tc_2C$, $Rh_2C$, *etc*. It has been reported that large-area and high-quality monolayer $Mo_2C$ has been fabricated by chemical vapor deposition (CVD).[39] Thus, in the following, we comparably study the 1T sandwich $Mg_2C$, $Be_2C$, Janus MgBeC, and $Mo_2C$.

Fig. 1(b, c) compare the $\kappa$ and the percentage contribution of each phonon branch of $Mg_2C$, Janus MgBeC, $Be_2C$, and $Mo_2C$. Surprisingly, it is found that the thermal transport properties of the 1T-sandwich structure are insensitive to the variation of components, which maintain a low thermal conductivity despite the substitution of different atoms. Specifically, the $\kappa$ of Janus MgBeC (8.26 W/mK) is lower than that of $Be_2C$ (14.80 W/mK)[40] but is higher than that of $Mg_2C$ (3.74 W/mK)[27]. The intriguing phenomenon is similar to the anomalous high $\kappa$ of the $Ga_{0.25}Al_{0.75}N$ alloy compared to GaN and AlN.[41] In addition, the $\kappa$ of $Mo_2C$ is only 5.13 W/mK. Thus, all the $\kappa$ of the four materials are extremely low (< 15 W/mK), which indicates the composition insensitive of the $\kappa$ of 1T-sandwich structure. The stable behavior promises its wide applications in lots of fields. For instance, impurities usually induce additional phonon scattering and lower the $\kappa$ of materials. With the stable behavior of thermal conductivity, the inevitable influence of impurities on the $\kappa$ of materials in practical applications and experiments can be effectively solved.

**3.2 Phonon dispersion**

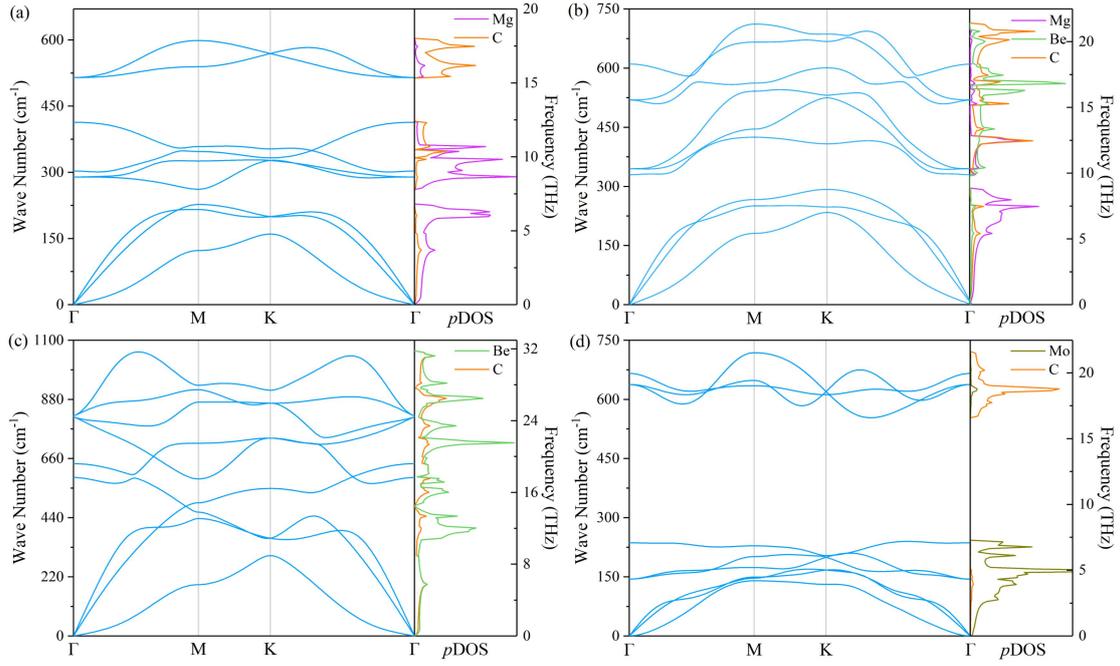

Fig. 2 The phonon dispersion and the corresponding atomic partial density of states (pDOS) for (a) $Mg_2C$, (b) Janus MgBeC, (c) $Be_2C$, and (d) $Mo_2C$.

The dynamic stability of the four materials is proved by the absence of imaginary frequency in the phonon dispersions, as shown in Fig. 2. Moreover, the thermal stability of the newly predicted Janus MgBeC is also confirmed based on the AIMD simulations (Supplemental Fig. S2). It is revealed that the highest frequency of optical branch of $Be_2C$ is up to 31.8THz, followed by Janus MgBeC (21.4THz) and $Mg_2C$ (18THz)[42], which is consistent with the sequence of average mass. It is worth noting that there are two band gaps in $Mg_2C$, with one between the acoustic and the optical branches and another inside the optical branches due to the great atomic mass difference. Based on the pDOS, it is confirmed that the lighter C atoms contribute the highest two optical branches. Specifically, the vibrations frequency of the C atoms in $Mo_2C$ is anomalous high (21.6THz), in good agreement with previous study[43], which leads to the widest band gap inside the optical branches. The reason lies in the largest atomic mass difference in $Mo_2C$ compared with $Mg_2C$, Janus MgBeC, and $Be_2C$. Thus, the band gap in the phonon spectrum decreases with the reduction of the atomic mass difference. As a result, the coupling between the high-frequency optical and acoustic phonon branches is the weakest in $Mo_2C$, increasing the contribution of the optical branches to the thermal conductivity, as shown in Fig. 1b. Besides, the independent ZA branch of $Mg_2C$, Janus MgBeC, and $Be_2C$ also significantly affect the contribution of each branch to the thermal conductivity.

## 3.3 Mode level analysis

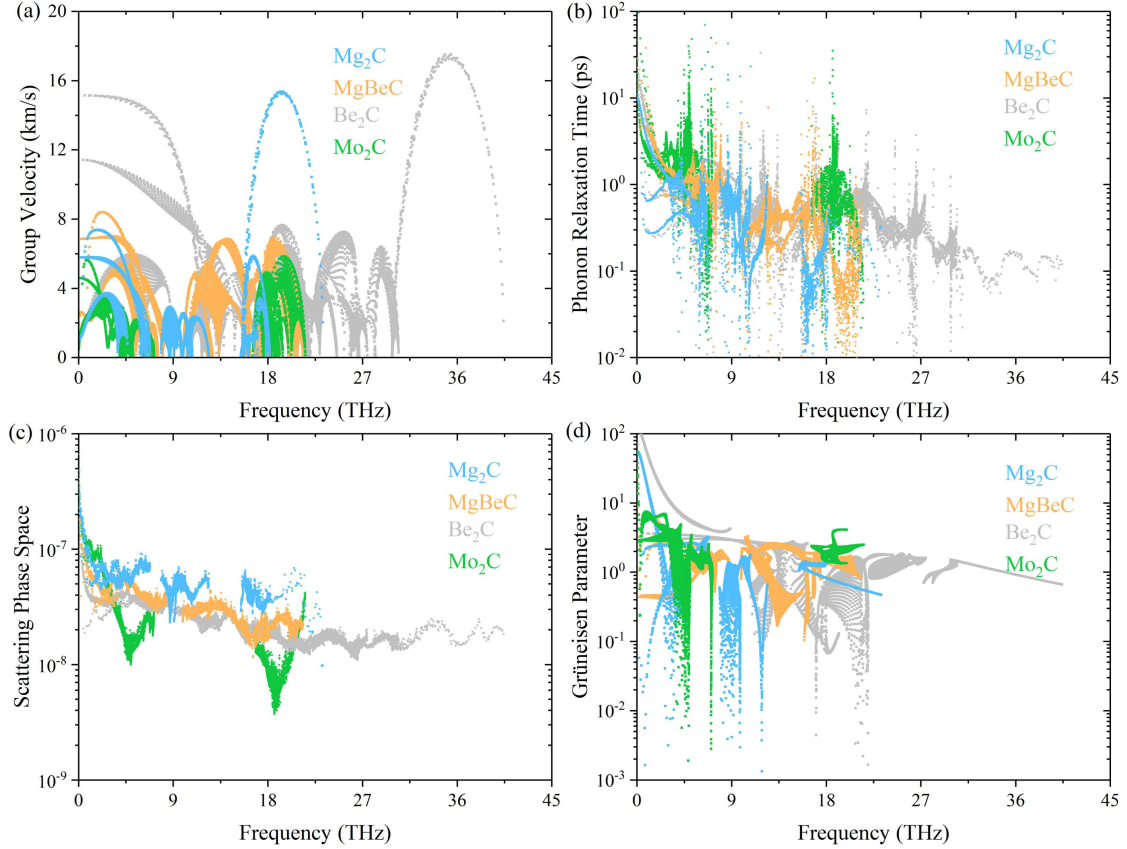

Fig. 3 Comparison of mode level (a) phonon group velocity, (b) phonon relaxation time, (c) total scattering phase space, and (d) absolute value of Grüneisen parameter as a function of frequency for $Mg_2C$, Janus MgBeC, $Be_2C$, and $Mo_2C$.

To make a deep understanding of the stable thermal conductivity in the 1T-sandwich structure, a comparative study is performed on the group velocity, phonon relaxation time, scattering phase space, and Grüneisen parameter, as shown in Fig. 3. It can be found that the group velocities of the four materials are precisely similar, concentrating in the range of 0-8 km/s (Fig. 3a). As reflected in the phonon spectrum, $Be_2C$ has the highest group velocity, followed by Janus MgBeC, which is consistent with the phenomena of thermal conductivity. However, it is worth noting that $Mo_2C$ has the lowest group velocity but not $Mg_2C$, mainly due to the larger atomic mass difference in $Mo_2C$. The frequency of optical branches contributed by Mo atoms is ultralow and the frequency of acoustic branches is the lowest in the four materials, which leads to the largest phonon band gap, flattest branches, and lowest group velocity.

Regarding phonon relaxation time (Fig. 3b), all the four materials are in the same magnitude, leading to the stable behavior of low thermal conductivity. However, there exists a slight difference among the $Mg_2C$, Janus MgBeC, $Be_2C$, and $Mo_2C$. The phonon relaxation time of $Mg_2C$ is the lowest, which is a critical reason for its lowest thermal conductivity. It is worth noting that the relaxation time of optical branches in $Mo_2C$ is higher than $Mg_2C$ and Janus MgBeC, which leads to a higher contribution percentage to the $\kappa$ (Fig. 1b). The scattering rate is negatively correlated with phonon relaxation time ($\Gamma = 1 / 2\tau$)[44], which can be divided into Absorption/Emission or Normal (N)/Umklapp (U) process as shown in Supplemental Fig. S3.[45] It can be found that the N process is relatively strong, leading to the strong phonon hydrodynamics and the large difference between the thermal conductivity predicted by RTA and iterative methods. In short, the total scattering rates of the four materials in each frequency range are about 10 $ps^{-1}$, which corresponds to the small relaxation time and low thermal conductivity.

The scattering rate is governed by both the scattering phase space[46] and the absolute value of the Grüneisen parameter. According to Fig. 3c and d, it is obvious that the slight differences in phonon relaxation time of $Mg_2C$, Janus MgBeC, $Be_2C$, and $Mo_2C$ are mainly caused by the scattering phase space, and the Grüneisen parameters are well consistent with each other. The relaxation time of $Mo_2C$ is the highest among the four materials because of the lowest scattering phase space. On the contrary, monolayer $Mg_2C$ has the highest scattering phase space and the lowest relaxation time. We further decompose the scattering phase space into the Emission/Absorption process to have a deep understanding. It can be seen from the Supplemental Fig. S4 that the absorption process is dominant in the low frequency, while the emission process is dominant in the high-frequency phonon mode.[47,48] Thus, the $Mo_2C$ has the smallest scattering phase space due to the widest band gap.

**3.4 Insight from electronic structures**

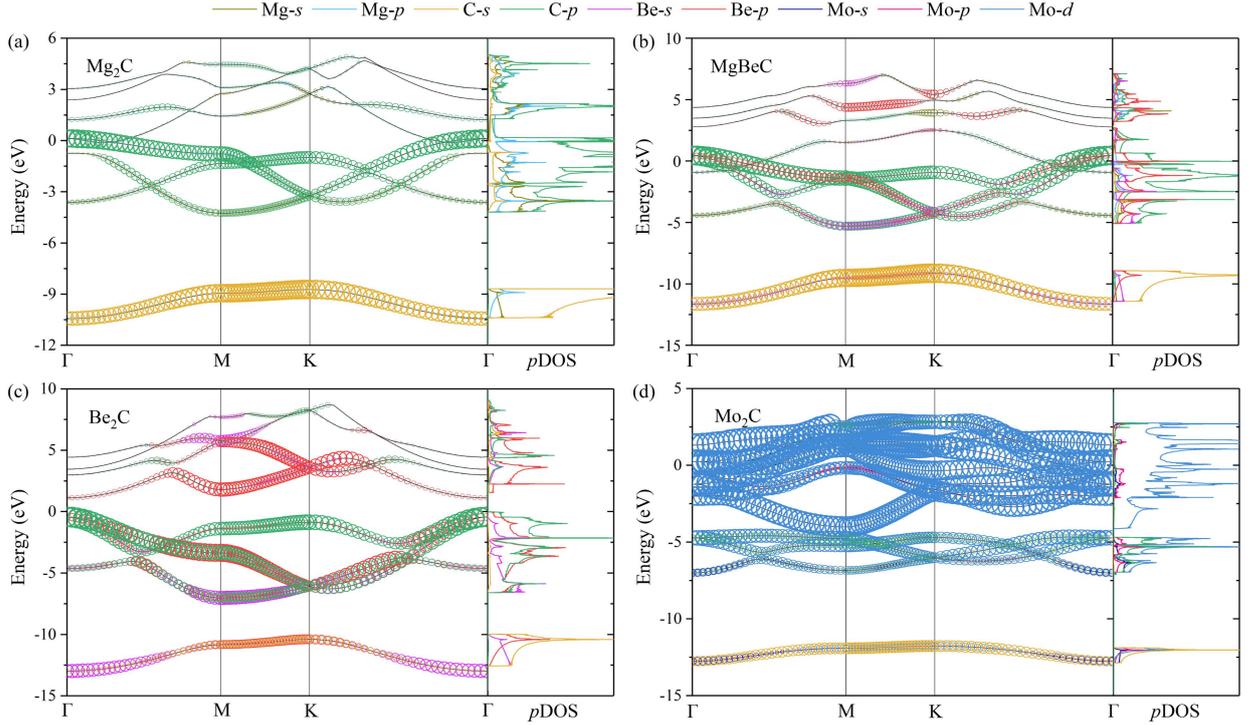

Fig. 4 Orbital projected electronic band structures and partial density of states (pDOS) of (a) Mg$_2$C, (b) Janus MgBeC, (c) Be$_2$C, and (d) Mo$_2$C.

To get a fundamental understanding on the mechanism from the microscopic perspective for the stable heat transfer performance of the 1T-sandwich structure with light C atoms in the middle plane, the electronic structures of monolayer Mg$_2$C, Janus MgBeC, Be$_2$C, Mo$_2$C were comparably analyzed, as shown in Fig. 4. The electronic band structures of Mg$_2$C, Janus MgBeC, and Be$_2$C are similar in the same 1T-sandwich structures, which are quite different from Mo$_2$C. There exists a large band gap of 1.55 eV in Be$_2$C. The band gap decreases with the relatively increasing atomic mass. Thus, the band gap can be purposely modulated by controlling the atom substitution, which may benefit its application in electronic, optoelectronic, and thermoelectric devices. Since the electronegativity of Mg (1.31) is smaller than that of Be (1.57), the electronegativity difference between C (2.55) and Mg is larger, and the polarity of Mg-C bonds is stronger. Mg and Be are the elements of group ⅡA with a similar valence electron configuration, while Mo is a transition metal element. As shown in Fig. 4d, the $d$ orbital electrons of Mo make a significant contribution to the valence band maximum (VBM) and conduction band minimum (CBM), resulting in the metallicity of Mo$_2$C. In contrast, the VBM is mainly contributed by C-$p$, Mg-$s$, or Be-$s$ orbitals in Mg$_2$C,

Janus MgBeC, and Be$_2$C. As there are only two valence electrons (*s*) of atoms from group Ⅱ A, and four valance electrons of C atoms, the resonant bonds are formed in Mg$_2$C, Janus MgBeC, and Be$_2$C as shown in Supplemental Fig. S5, which is consistent with the analysis drawn in our previous study.[27] Thus, the thermal conductivity is intrinsically low due to the resonant bonding. A similar electronic structure contributes to the consistent phenomenon of anharmonic vibration (Grüneisen parameter), phonon relaxation time, and group velocity. Therefore, through the component design based on the 1T-sandwich structure, the apparent electronic band design can achieve the stable behavior of low thermal conductivity, which would be beneficial to the design and manufacturing of electronic devices.

## 3.5 Size effect

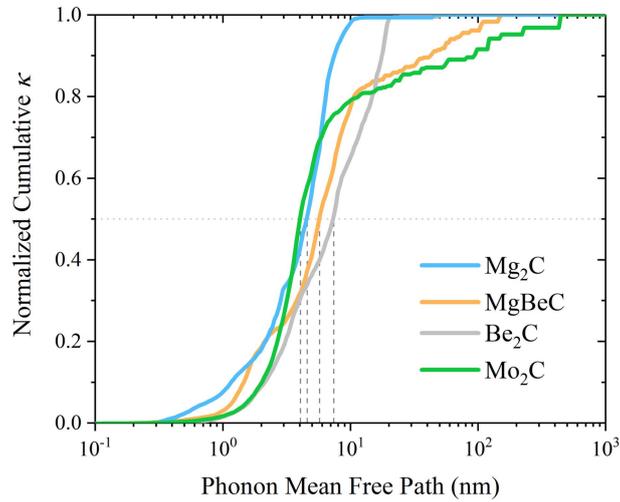

Fig. 5 Comparison of normalized cumulative $\kappa$ with respect to phonon mean free path (MFP) for Mg$_2$C, Janus MgBeC, Be$_2$C, and Mo$_2$C. The representative MFP is marked onsite corresponding to the 50% $\kappa$.

For practical device applications, the size would significantly affect the quality factors related to heat transfer performance.[49,50] When the characteristic size is smaller than the phonon means free path, the ballistic transport is dominant, and the effective lattice $\kappa$ will be limited to be smaller than the intrinsic value. It can be seen that the difference of rMFP (corresponding to 50% $\kappa$) among Mg$_2$C, Janus MgBeC, Be$_2$C, and Mo$_2$C is minimal, which are around 4-7.2 nm. When the size reaches $10^3$ nm, the $\kappa$ of the four materials converges. Such behaviour means that the MFP is lower than most materials that have been extensively studied, such as graphene[51], phosphorene[52], *etc*. For instance, the MFP of $Mg_2C$ is less than 10 nm, which benefits $Mg_2C$ to be fabricated as ultra-thin devices. By comparison, it is found that when the $\kappa$ reaches 0.8 of the converged $\kappa$, the $\kappa$ of Janus MgBeC and $Mo_2C$ increases slower than $Mg_2C$ and $Be_2C$. Therefore, finite-size has a more significant effect on the $\kappa$ of Janus MgBeC and $Mo_2C$ compared to $Mg_2C$ and $Be_2C$.

## 4. Conclusion

Based on the state-of-art first-principles calculations, we solved the phonon BTE and carried out a comparative study on the heat transport performance of $Mg_2C$, Janus MgBeC, $Be_2C$, and $Mo_2C$. The series of novel 2D materials are constructed through component design based on the framework of the 1T-sandwich structure, where the novel Janus MgBeC was predicted by replacing the two sides with Be and Mg atoms, respectively. The 1T-sandwich structure is the same as the well-known transition metal dichalcogenide (TMD) but with light Carbon atoms in the middle plane. Stable behaviour of the low $\kappa$ is found in the 1T-sandwich structure with the light Carbon atoms in the middle layer. The $\kappa$ of $Mg_2C$, Janus MgBeC, $Be_2C$, and $Mo_2C$ are calculated to be 3.74, 8.26, 14.80, 5.13 W/mK, respectively. The phonon group velocity, relaxation time, and rMFP of the four materials are on the same order of magnitude, which explains the slight variation of $\kappa$. From the analysis of the electron structure, the consistent structure of the resonance bond is responsible for their similar performance and the stable behaviour of low $\kappa$. The stably low $\kappa$ of the 1T-sandwich structure indicates the insensitivity of thermal transport to the component, which reveals the primary effect of geometry structure. Thus, impurities have a weak performance to materials with the 1T-sandwich structure, which provides an effective way for device fabrication and paves the way for the design and application in the thermoelectric, and thermal protection, *etc*.

## Acknowledgments

This work is supported by the National Natural Science Foundation of China (Grant No. 52006057), the Fundamental Research Funds for the Central Universities (Grant Nos. 531118010471, 531119200237, and 541109010001). The numerical calculations in this paper have been done on the supercomputing system of the National Supercomputing Center in Changsha. The Lichtenberg high performance computer of the TU Darmstadt is gratefully acknowledged for the computational resources.

# The stable behavior of low thermal conductivity in 1T-sandwich structure with component design

E Zhou[1], Jing Wu[1], Chen Shen[2], Hongbin Zhang[2], Guangzhao Qin[1,−]

*[1]State Key Laboratory of Advanced Design and Manufacturing for Vehicle Body, College of Mechanical and Vehicle Engineering, Hunan University, Changsha 410082, P. R. China*

*[2]Institut für Materialwissenschaft, Technische Universität Darmstadt, Darmstadt, 64289, Germany*

−  Author to whom all correspondence should be addressed. E-Mail: gzqin@hnu.edu.cn

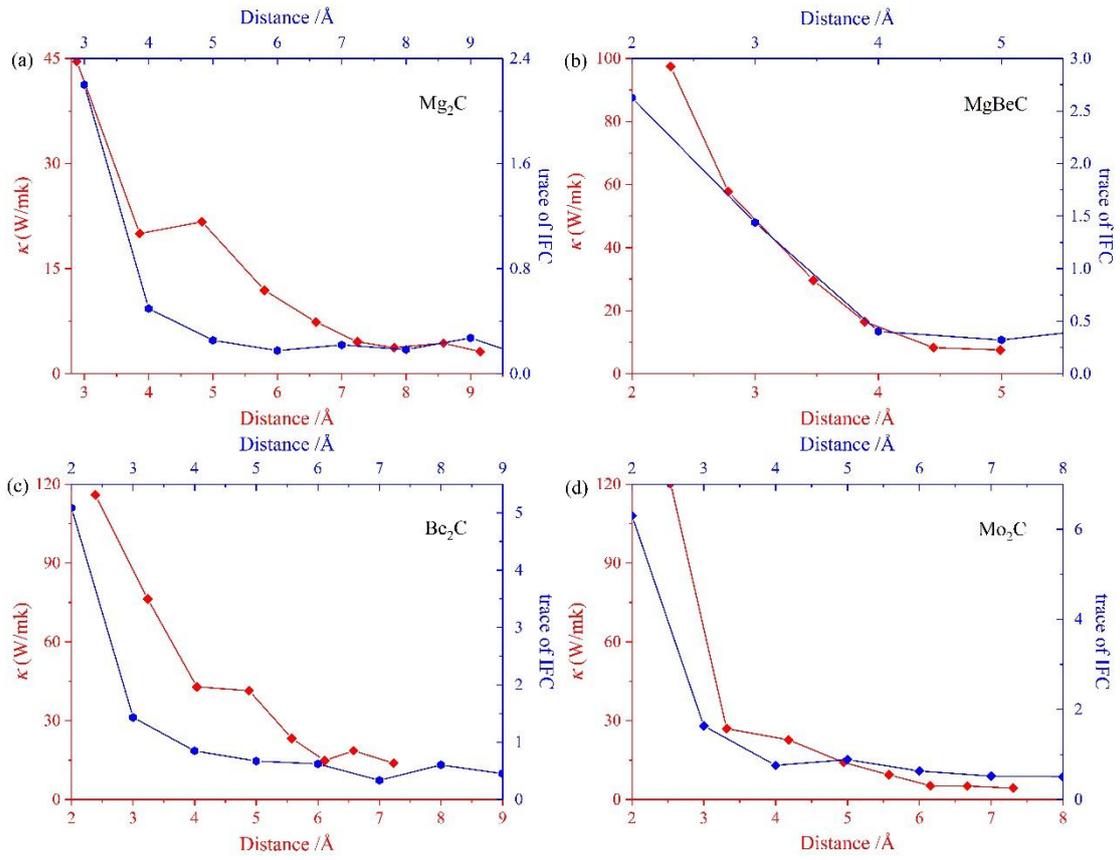

Supplemental Fig. S1. The thermal conductivity and the root mean square of second order IFC tensor with different cut-off distance of $Mg_2C$, $MgBeC$, $Be_2C$, $Mo_2C$. The red (blue) axis corresponds to the red (blue) curve.

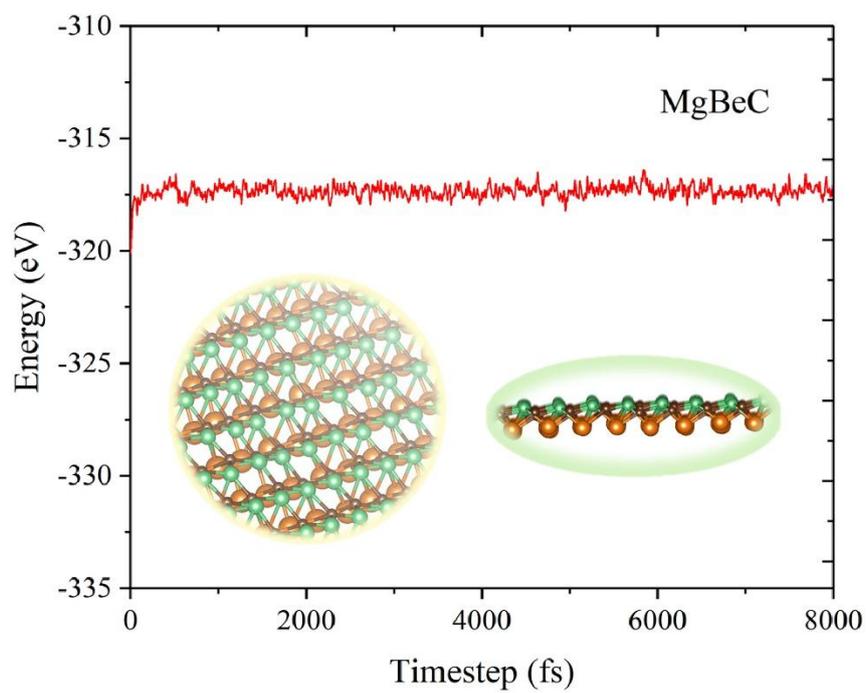

Supplemental Fig. S2. Structures and thermal stability of MgBeC from the ab-initio molecular dynamics simulation at 300 K. A timestep is 1fs.

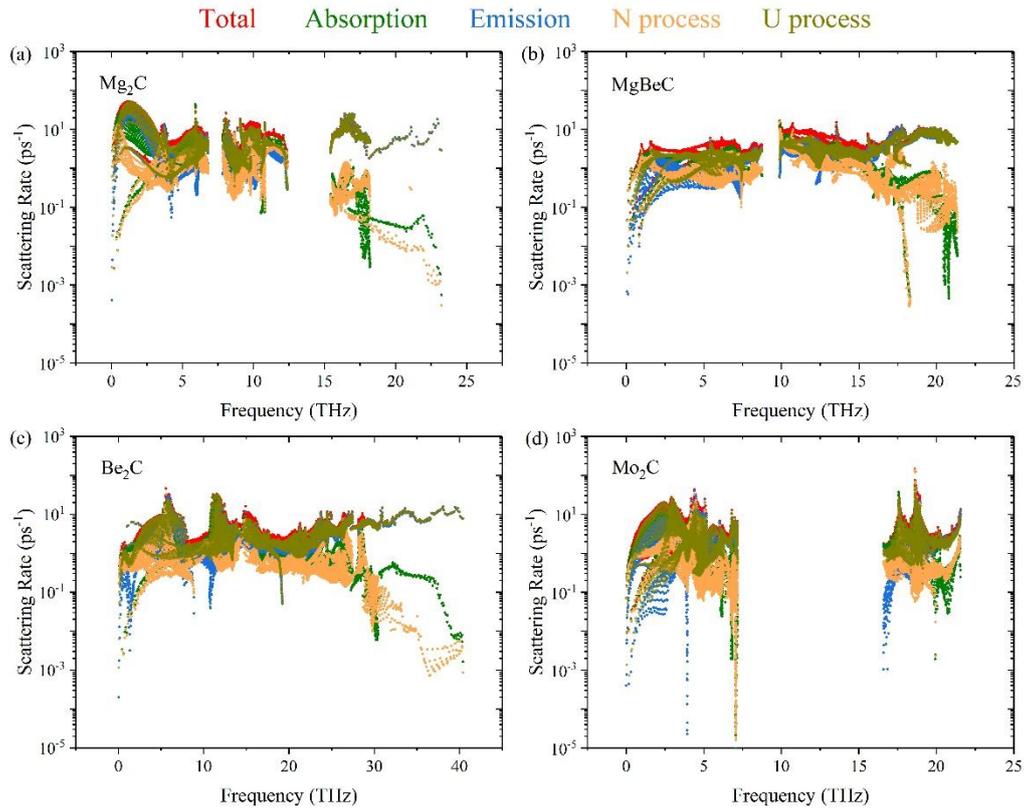

Supplemental Fig. S3. The phonon-phonon scattering rate as a function of frequency for Mg$_2$C, MgBeC, Be$_2$C, and Mo$_2$C at 300 K. Red, green, blue, orange, and brown points represent the scattering rate in total, absorption, emission, Normal, and Umklapp processes, respectively.

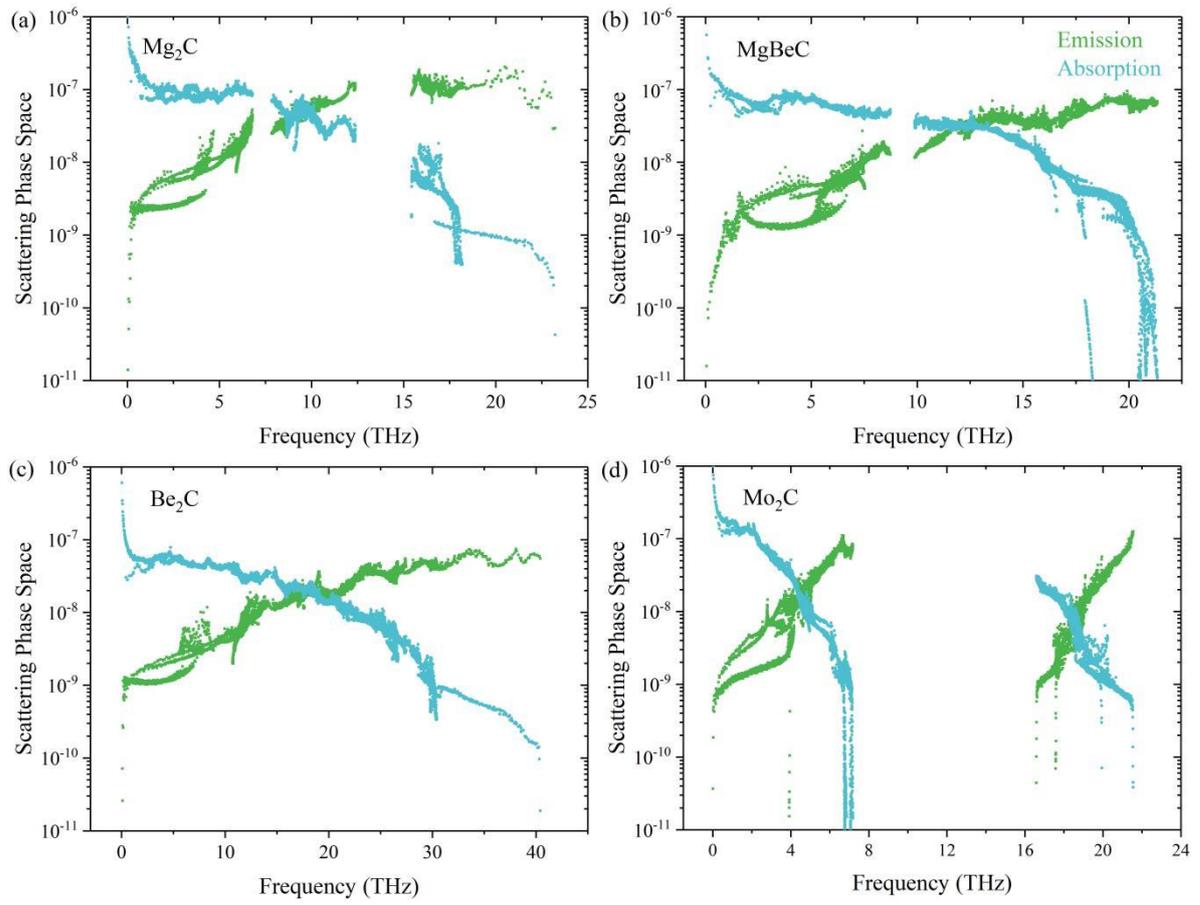

Supplemental Fig. S4. The scattering phase space classified into Absorption/Emission processes of (a) $Mg_2C$, (b) $MgBeC$, (c) $Be_2C$, and (d) $Mo_2C$.

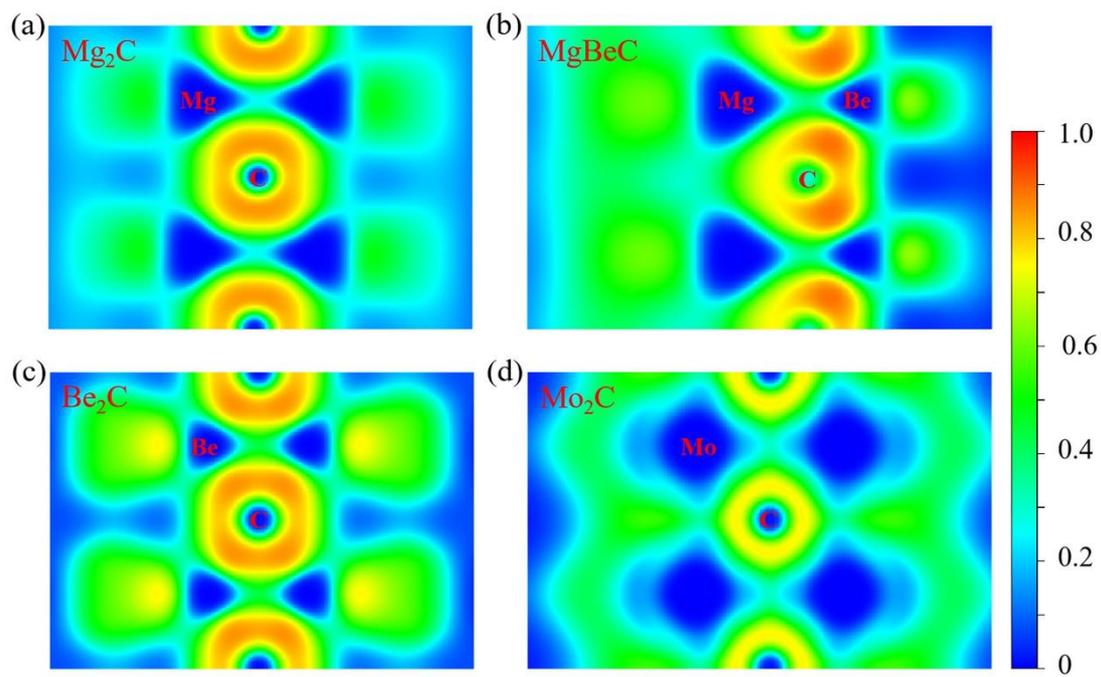

Supplemental Fig. S5. The electron local function (ELF) of (a) $Mg_2C$, (b) MgBeC, (c) $Be_2C$, and (d) $Mo_2C$.